# On-chip excitation of single germanium-vacancies in nanodiamonds embedded in plasmonic waveguides

Hamidreza Siampour[*,1], Shailesh Kumar[1], Valery A. Davydov[2], Liudmila F. Kulikova[2], Viatcheslav N. Agafonov[3], and Sergey I. Bozhevolnyi[1]

[1] Centre for Nano Optics, University of Southern Denmark, Campusvej 55, Odense M, DK-5230, Denmark

[2] L.F. Vereshchagin Institute for High Pressure Physics, Russian Academy of Sciences, Troitsk, Moscow, 142190, Russia

[3] GREMAN, UMR CNRS CEA 6157, Université de Tours, 37200 Tours, France

[*]E-mail: hasa@mci.sdu.dk

**Abstract-** Monolithic integration of quantum emitters in nanoscale plasmonic circuitry requires low-loss plasmonic configurations capable of confining light well below the diffraction limit. We demonstrate on-chip remote excitation of nanodiamond-embedded single quantum emitters by plasmonic modes of dielectric ridges atop colloidal silver crystals. The nanodiamonds are produced to incorporate single germanium-vacancy (GeV) centers, providing bright, spectrally narrow and stable single-photon sources suitable for highly integrated circuits. Using electron-beam lithography with hydrogen silsesquioxane (HSQ) resist, dielectric-loaded surface plasmon polariton waveguides (DLSPPWs) are fabricated on single crystalline silver plates so as to contain those of spin-casted nanodiamonds that are found to feature appropriate single GeV centers. The low-loss plasmonic configuration enabled the 532 nm pump laser light to propagate on-chip in the DLSPPW and reach to an embedded nanodiamond where a single GeV center is incorporated. The remote GeV emitter is thereby excited and coupled to spatially confined DLSPPW modes with an outstanding figure-of-merit of 180 due to a ~6-fold Purcell enhancement, ~56% coupling efficiency and ~33 μm transmission length, revealing the potential of our approach for on-chip realization of nanoscale functional quantum devices.

## INTRODUCTION

There have been developed different metal-dielectric configurations that support propagating plasmonic modes confined beyond the diffraction limit allowing for strong light-matter interaction down to the single-photon level[1-5]. Various types of surface plasmon polariton (SPP) based structures, such as metal nanowires (NW)[6-9], parallel NWs[10], V-grooves (VGs)[11], and wedge waveguides[12], have been demonstrated to guide single plasmons, quanta of propagating plasmonic modes, for potential quantum applications. However, the practical realization of SPP-based integrated quantum photonics has remained elusive due to several formidable challenges, including the implementation of functional components within the limitations arising from inevitable SPP propagation losses and efficient coupling of SPP mode to the quantum emitter. Recently, by using relatively low-loss dielectric-loaded SPP waveguides (DLSPPWs) structured on silver (Ag) film, simple quantum plasmonic circuits composed of embedded nanodiamonds with nitrogen-vacancy (NV) centers have been demonstrated[13,14]. The nanodiamonds hosting color centers with their photostable single photon emission and optically readable spin states are promising candidates to build integrated quantum devices, for example integration into plasmonic circuits. In addition to NV center, a family of diamond color centers based on group-IV elements in the periodic table, i.e., silicon-vacancy (SiV)[15-21], germanium-vacancy (GeV)[22-26], and tin-vacancy (SnV) centers[27,28], have attracted attention due to their structural symmetries leading to high emission into zero-phonon lines (ZPLs), accompanied by bright, spectrally narrow emission lines and indistinguishability of photons emitted from different emitters. The SiV centers exhibit an optical transition at longer wavelength (ZPL at 738 nm) operating with smaller SPP loss in the metal[29]. At the same time, the SiV exited state decay is dominated by the nonradiative relaxation, causing lower quantum efficiency for SiV centers[30]. More recently, the GeV centers in diamond have shown stronger coupling between emitters and photons than SiV due to their higher quantum efficiency and larger absorption cross section[23]. These properties of GeV center can be utilized to facilitate remote excitation of GeV centers incorporated in a plasmonic chip.

In this work, we demonstrate on-chip generation and propagation of spectrally narrow single optical plasmons excited by GeV centers in nanodiamonds using DLSPPWs. The nanodiamonds are produced using high-pressure high-temperature (HPHT) diamond synthesis method, and Ge is introduced during the growth process to incorporate single GeV centers (see details in Methods section). Furthermore, we propose and demonstrate a hybrid approach using DLSPPWs structured on single Ag crystals that feature considerably lower SPP damping rates in comparison with Ag films fabricated by other techniques[13,31]. The latter allows us to realize sufficiently long SPP propagation at both the excitation and emission wavelengths of GeV centers, facilitating thereby the remote excitation of GeV centers in nanodiamonds incorporated in a plasmonic chip. The colloidal Ag crystal flakes have been synthesized using a recently reported recipe based on platinum (Pt)-nanoparticle-catalyzed and ammonium hydroxide ($NH_4OH$)-controlled polyol reduction method[31]. Figure 1 shows a schematic of the device layout, introducing the working principle of on-chip remote excitation. The hybrid plasmonic configuration enabled the green laser light to propagate on-chip, inside the DLSPPW and reach to an embedded nanodiamond containing a single GeV center. The remote GeV emitter is thereby excited and single photons are channeled to a DLSPPW mode. Finally, the ability of GeV-DLSPPW coupled system to achieve efficient long-range energy transfer is compared with other hybrid quantum plasmonic systems, revealing an exceptional figure of merit (FOM) of 180 due to a ~6-fold Purcell enhancement, ~56% coupling efficiency and ~33 μm transmission length at λ=602 nm (ZPL, GeV).

In the first designed experiment, a silicon wafer is coated with a Ag film of 250 nm thickness (*via* thermal evaporation in a vacuum pressure of $2 \times 10^{-7}$ Torr, and a rate of 40 Å.s$^{-1}$), on which gold markers are made, and subsequently the synthesized GeV nanodiamonds are spin coated. A 1-nm layer of Poly(allylamine hydrochloride) (PAH) is put on the Ag layer to improve distribution and attachment of the nanodiamonds to the Ag surface[32]. The sample is then characterized by scanning in a fluorescence confocal microscope. Lifetime, spectrum, and autocorrelation measurements are taken for the nanodiamonds on the Ag film, and the position of nanodiamonds containing single GeV centers are determined with respect to the gold markers. Hydrogen silsesquioxane (HSQ) e-beam resist (Dow Corning XR-1541−006) is then spin coated (1200 rpm, 1 min) to make a 180 nm film on the Ag-coated substrate. Waveguide is fabricated, using electron beam lithography, onto the nanodiamond that is found to be a single photon emitter. Schematic of the device layout for on-chip direct excitation of nanodiamonds is illustrated in Figure 2a and Figure 2b. Figure 2c shows a distribution of normalized $|E|^2$

(where E is the electric field) of the DLSPPW mode with a cross-section of 180 nm height and 250 nm width of HSQ (dielectric constant of 1.41) on Ag surface[33] at λ=602 nm. An atomic force microscope (AFM) image of the fabricated waveguide is presented in Figure 2d (left). Postfabrication, when the GeV center is directly excited with green pump laser, charge coupled device (CCD) camera image shows three spots, ND, A and B (Figure 2d, right). This shows excitation and emission of the GeV emitter (ND), coupling of GeV emission to the DLSPPW mode, propagation and subsequent emission from the gratings at the two ends (A and B). The emission spectra of the GeV center taken before (Figure 2e) and after coupling (Figure 2f) indicate a ~3 times enhancement in the emission rate of the GeV center after coupling to the confined DLSPPW mode (for the same excitation power). Figure 2g shows the spectra from the out-coupling grating couplers A and B, indicating the on-chip transmission of spectrally narrow surface plasmon mode excited by GeV emitter. Antibunching dip observed in the second-order correlation function of the GeV center both before (Figure 2h) and after (Figure 2i) fabrication of the waveguide, shows a single photon emission ($g^2(\tau = 0) < 0.5$). The measured data is fitted with single exponential model[23]. In Figure 2j, lifetime of the GeV center before (grey) and after (red) coupling is presented. A lifetime reduction (from ~12.3 ns to ~3.8ns) is observed for the coupled GeV center. The lifetimes are obtained by tail fitting of the measured data with a single exponential. First few ns of the data is avoided in the fitting, as it arises from background fluorescence. We have observed lifetime reduction changes in the range of 2.5 to 3.5, and on average, the lifetime decreased by a factor of ~3 which is in addition of the 2 times reduction due to a Ag plane surface (Supplementary S2), giving a ~6-fold Purcell enhancement in the total decay rate.

We have estimated propagation length of the DLSPPW on Ag film by comparing the attenuation of fluorescence signals $P_A$ and $P_B$ at the two ends of the waveguide and the corresponding propagation distances $L_A$ and $L_B$ from coupled GeV center. The 1/*e* propagation length, $L_P$, is extracted from the fluorescence signals at the two ends using $P_A/P_B = \exp[(L_A - L_B)/L_P]$, where we assume symmetric coupling in two directions, uniform losses across the waveguide and the same out-coupling efficiency at the grating ends. The collected data are fitted to obtain the propagation length of 9 ± 3 μm for the GeV-DLSPPW hybrid system on Ag film which is smaller than NV- DLSPPW system[13] because of the higher material loss in the metal at lower wavelength region corresponding to emission from GeV centers.

Using single-crystalline Ag flake instead of Ag film, significantly enhanced the DLSPPW propagation length (Figure 3). Figure 3a shows an SEM image of a fabricated HSQ waveguide on a Ag crystal flake. Optical characterization of the waveguide shows transmission of green laser light (wavelength: 532nm) through DLSPPW mode for the polarization parallel to the waveguide axis (Figure 3b). We have measured the transmission for several waveguides with different lengths, achieving an extraordinary long propagation length of ~11.8 μm for the green laser light through the low-loss DLSPPW (see Supplementary S3 for details).

In the second designed experiment, we employ the capability of green light transmission in the DLSPPW on Ag crystals and demonstrate remote excitation of GeV nanodiamonds embedded in low-loss DLSPPWs. We spin-coat a small amount of the synthesized solution of Ag crystal flakes on an Ag-coated silicon wafer. A PAH layer is put on the Ag film to improve sticking of the Ag flakes to the substrate[32]. Gold markers are made on Ag flakes, and subsequently the nanodiamonds with GeV centers are spin-coated. The sample is then characterized by scanning in a fluorescence confocal microscope. Spectrum, and autocorrelation measurements are taken for the nanodiamonds on Ag crystals. The HSQ e-beam resist is then spin coated on the substrate and waveguide is fabricated onto the nanodiamond, which is found to contain a single GeV center using electron beam lithography. Schematic of the device layout and working principle for on-chip remote excitation is illustrated in Figure 4a. Figure 4b shows an AFM image of the fabricated waveguide (left), and a galvanometric mirror scan image (right) when the waveguide is excited from end B with green pump laser and a fluorescence image of the focal plane is taken. Emission from the embedded GeV nanodiamond located inside the waveguide confirms a remote excitation of GeV center, and emission from the gratings at the two ends of the waveguide indicate the coupling of GeV center to the DLSPPW mode. Figure 4c shows the emission spectrum of the GeV center taken before coupling (nanodiamond on Ag crystal). The emission spectra after coupling for the remote excited GeV center (solid line). Spectrum for the same GeV center when excited directly (dotted line) is also presented in Figure 4d. CCD images for the coupled system when excited directly and with a linear polarizer placed in the detection path are presented in Figure 4e. A comparison of the two images in Figure 4e, clearly indicates a strongly polarized emission from the end of waveguide due to the coupling of emission to the fundamental transverse magnetic (TM) mode of the DLSPP waveguide, and propagation and subsequent scattering from the ends. Figure 4f shows the spectrum measured at the grating end A, when GeV center is remotely excited (Figure 4b). In Figure 4g, we present the second-order correlation function for the GeV center, confirming single photon emission ($g^2(\tau=0) < 0.5$).

We calculate the GeV emitter's decay rate into the DLSPPW mode using two dimensional (2D) finite-element modeling (FEM) method[34]. GeV center is modelled as a single dipole emitting at 602nm in the simulations. Up to 4-fold decay rate to the plasmonic mode is predicted for a GeV center in waveguide, compared to its emission in vacuum. Figure 5a shows the plasmonic decay rate dependence for the optimum orientation of dipole on its position in the cross-section of the waveguide. Emission efficiency ($\beta$ factor) of the emitter to the DLSPPW mode is defined as the fraction of the emitted energy that is coupled to the plasmonic mode, i.e., $\beta = \Gamma_{pl}/\Gamma_{tot}$, where $\Gamma_{tot}$ is the total decay rate, including radiative decay rate, nonradiative decay rate, and the plasmonic decay rate ($\Gamma_{pl}$). A three-dimensional (3D) FEM model is implemented using COMSOL Multiphysics software and the total decay rate is extracted from the total power dissipation of the coupled emitter, as explained in refs [13,34,35]. The $\beta$ factor for a y- oriented GeV quantum emitter with a DLSPPW is simulated, as a function of position in the cross-section (Figure 5b). We used Palik's handbook of optical constants[33] for modeling of Ag refractive index. For optimum distance of the GeV emitter to the Ag surface, the $\beta$ factor can reach 62%. We measure the apparent plasmonic coupling factor, $\beta_{pl}$, using $\beta_{pl} \simeq (I_A + I_B)/(I_A + I_B + I_{GeV})$, where $I_A$ and $I_B$ are the emission intensity collected by the CCD camera at the ends of the waveguide and $I_{GeV}$ is the emission intensity measured at the GeV center position (i.e., free space emission not coupled to the DLSPPW mode). Here, we have assumed same collection efficiency from the GeV center as well as that from the ends, and have ignored the nonradiative decay. An apparent $\beta$ factor of 56% is found for the coupled DLSPP waveguide shown in Figure 4. This is in good agreement with the simulation result and predicts a satisfactory precision in deterministic positioning of the DLSPP waveguide onto the GeV nanodiamond. We have measured the 1/e propagation length of the GeV emission along the DLSPPW on Ag crystal using the same way as described for Ag film and obtain a value of 33±3 μm (see Supplementary S4). This is significantly larger than DLSPPW on polycrystalline Ag film and even higher than the propagation length of NV-DLSPPW[13], indicating a low material loss for the single crystalline Ag flakes. The ability of such system to achieve efficient long range energy transfer can be quantified by a figure of merit (FOM) defined as the product of the propagation length, Purcell factor ($\Gamma_{tot}/\Gamma_0$) and $\beta$-factor, normalized by the operational wavelength, as proposed in ref [11]. The GeV-DLSPPW coupled system reaches a FOM value of 180 ±25 ($\Gamma_{tot}/\Gamma_0$=6±1, $\beta$=0.56±0.03, $L_P$=33±3 µm, and $\lambda$=602 nm) for FOM, clearly outperforming previous demonstrations of quantum emitter-plasmonic waveguide (QE-PW) coupled systems.[6,10-13]. A careful comparison of GeV-DLSPPW on Ag crystal with other hybrid systems of QE-PW is presented in Figure 5c.

## CONCLUSIONS

We have demonstrated on-chip generation and transmission of spectrally narrow single optical plasmons excited by GeV nanodiamonds embedded in DLSPPWs. The extraordinary long propagation length for the green pump laser has been achieved with DLSPPWs on Ag crystals, enabling thereby the remote excitation of GeV centers through propagating DLSPPW mode in a plasmonic chip. The performance of the GeV-DLSPPW coupled system to achieve efficient long-range energy transfer has been quantified with the FOM of 180 by a ~6-fold Purcell enhancement, ~56% coupling efficiency and ~33 µm transmission length, that indicates a superior performance in comparison with the previous demonstrated systems. Our demonstration paves the way for integration of excitation laser, quantum emitter and plasmonic circuit on the same chip. Detection of single plasmons and two-plasmon interference has already been demonstrated on a chip[36,37]. With a combination of all these technologies, it will be possible, in the near future, to have all the elements of a quantum plasmonic circuit integrated on a chip.

## METHODS

**GeV nanodiamond synthesis-** High pressure – high temperature (HPHT) synthesis of nano-size diamonds with germanium-vacancy (GeV) color centers has been realized based on the hydrocarbon metal catalyst-free growth system. Tetraphenylgermanium $C_{24}H_{20}Ge$ (Sigma-Aldrich) has been used as the initial germanium-containing hydrocarbon compound. The synthesis was performed in a high-pressure apparatus of "Toroid" type. Cylindrical samples of the initial material (5 mm diameter and 3 mm height) obtained by cold pressing were put into graphite containers, which also served as a heater for the high-pressure apparatus. The experimental procedure consisted in loading the apparatus up to 8 GPa, heating up to the synthesis temperature and short isothermal exposure under constant load for 1-5 s. The obtained diamond products are then isolated by quenching to room temperature under pressure. The recovered samples have been characterized by X-ray diffraction, Raman spectroscopy, scanning (SEM) and transmission (TEM) electron microscopies. The results of such characterization of the obtained products, which are mixtures of ultranano-, nano- and submicrometer-size fraction of diamond, show that the formation of diamond occurs with virtually 100 % yield. Size-fractional separation of diamonds was carried out in several stages that consisted of ultrasonic dispersing of the diamond particles using UP200Ht dispersant (Hielscher Ultrasonic Technology), chemical treatment of the samples in mixture of three acids ($HNO_3$-$HClO_4$-$H_2SO_4$), and subsequent centrifugation of aqueous or alcohol dispersion of diamond powders.

## ACKNOWLEDGMENTS

This work was supported by the European Research Council, Advanced Grant 341054 (PLAQNAP). VAD and LFK thank the Russian Foundation for Basic Research (Grant No. 18-03-00936) for financial support.

## CONFLICT OF INTEREST

The authors declare no conflict of interest.

**FIGURES**

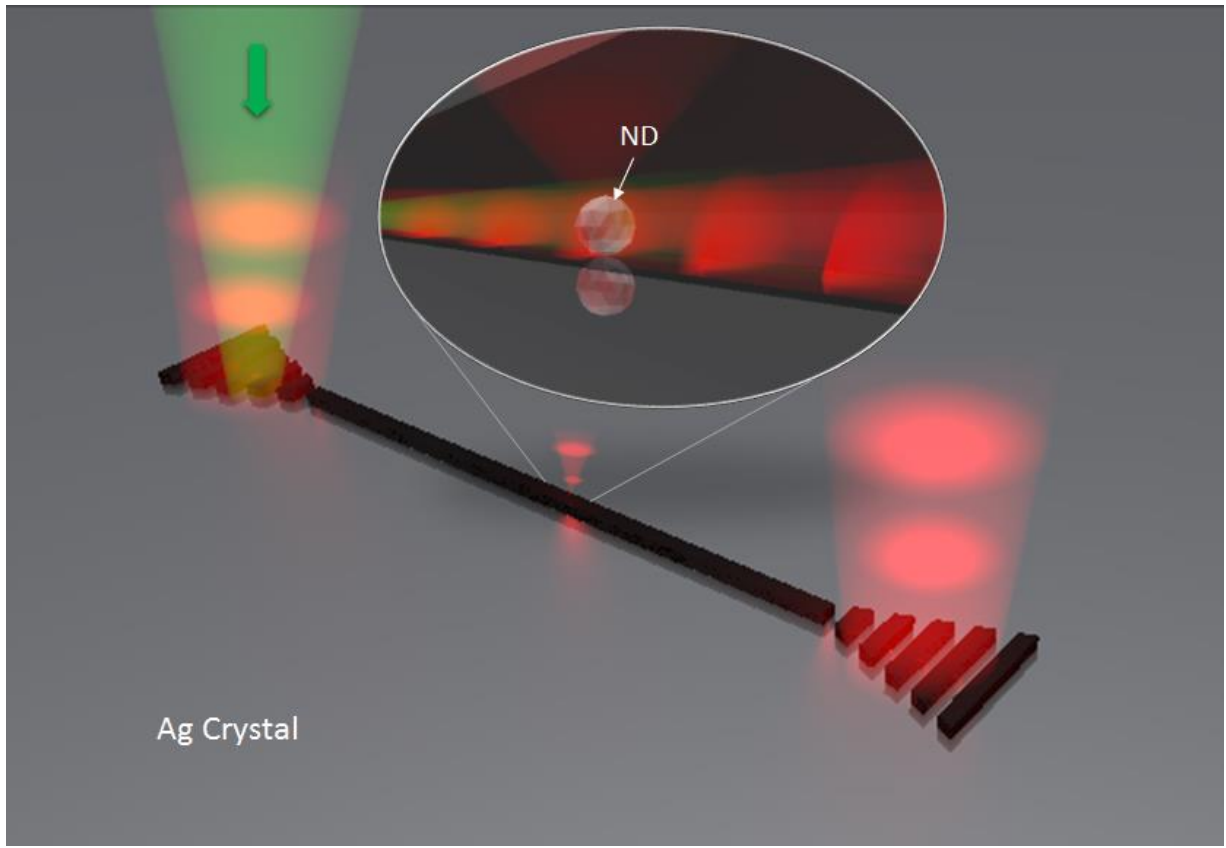

**Figure 1** Schematic of the device layout and working principle for on-chip excitation of a nanodiamond (ND) carrying spectrally narrow single GeV quantum emitters embedded in a DLSPP waveguide. A 532 nm pump laser light is coupled with a grating, and propagate on-chip in the low loss DLSPPW and reach to an embedded nanodiamond which contain a single GeV center. The remote GeV emitter is thereby excited, generating single optical plasmons propagating along the waveguide and outcoupling from the ends.

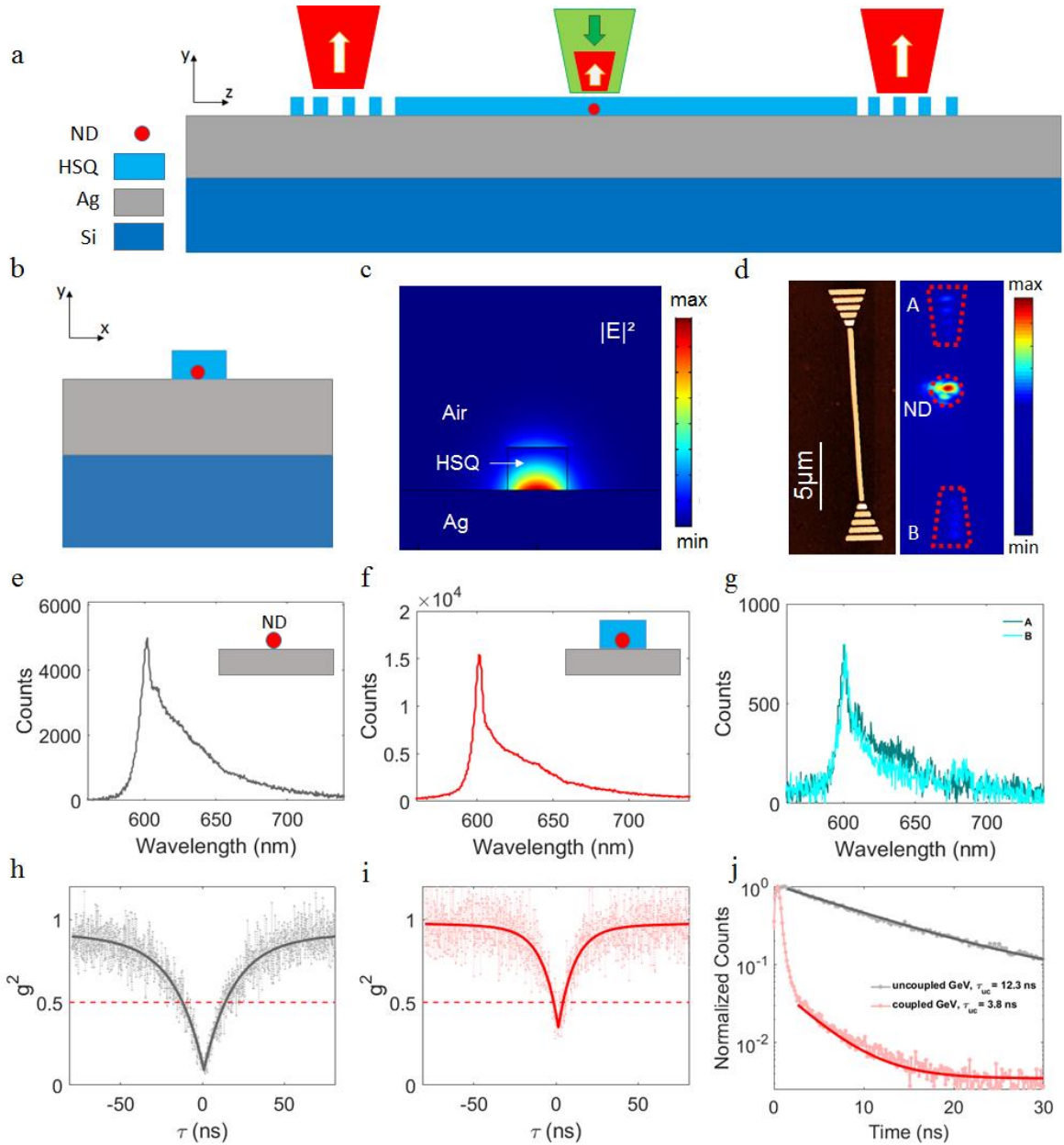

**Figure 2** (**a**, **b**) Schematic of sample layout and working principle of direct excitation of a GeV nanodiamond embedded in a plasmonic waveguide. (**c**) Simulated mode profile for the DLSPP waveguide at λ=602 nm (GeV ZPL). (**d**) AFM image of the fabricated waveguide (left), and CCD camera image of the whole structure where the nanodiamond is excited and a fluorescence image of the focal plane is taken (right). (**e-g**) Spectrum taken from uncoupled GeV (**e**), coupled GeV (**f**), and outcoupled light through the grating ends A and B (**g**). (**h**, **i**) Second order correlation of the GeV center before (**h**) and after (**i**) coupling to the waveguide. (**j**) Lifetime of the GeV center taken before (grey) and after (red) coupling.

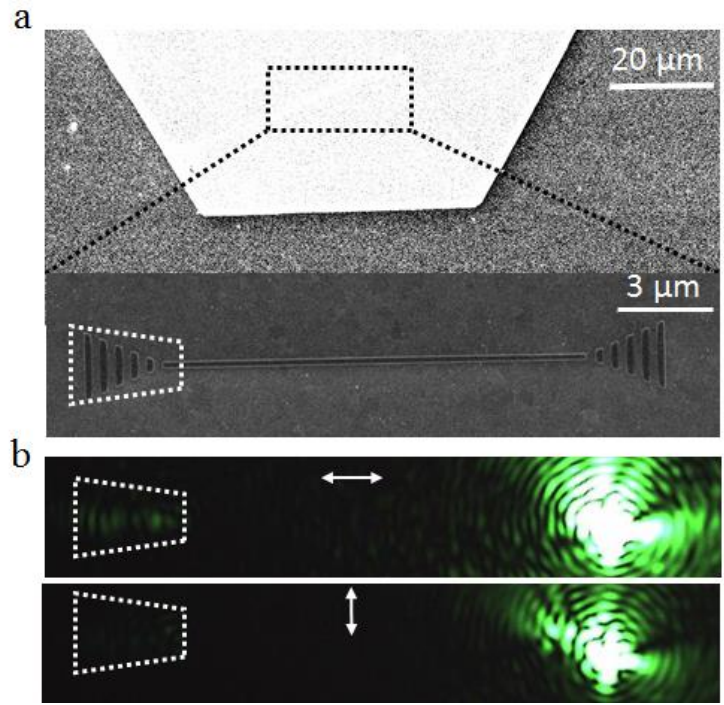

**Figure 3** Transmission of the 532 nm pump laser light along the low-loss plasmonic waveguide. (**a**) SEM image of a single crystalline Ag flake (top), and fabricated DLSPP waveguide on the Ag plate. (**b**) Optical characterization of the waveguide for parallel (top) and perpendicular (bottom) polarizations of 532 nm laser light.

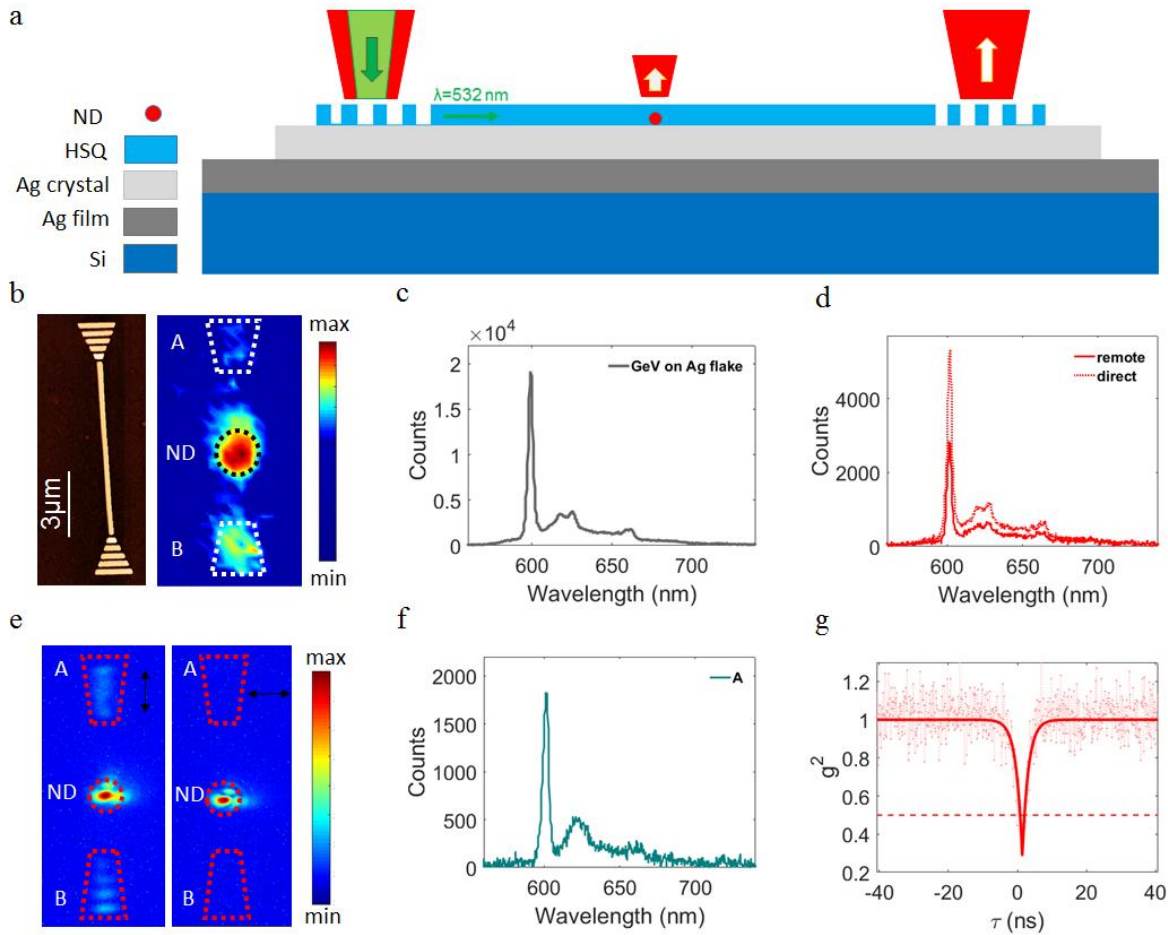

**Figure 4** (**a**) Schematic of sample layout for on-chip remote excitation of a GeV nanodiamond embedded in the plasmonic structure. (**b**) AFM image of fabricated waveguide (b, left) and galvanometric mirror scan image showing a remote excitation of the embedded GeV where the pump laser light is illuminated at the end B (b, right). (**c**, **d**) Spectra taken from the uncoupled GeV i.e., the nanodiamond on Ag plate (c) and from coupled GeV when excited remotely (d, solid line) and in the case of direct excitation (d, dotted line). (**e**) CCD images for the coupled system when excited directly and with a linear polarizer placed in the detection path are presented for two orthogonal polarizations, parallel (left) and perpendicular (right) to the waveguide axis. (**f**) Spectrum taken from outcoupled light through the grating end A in the case of remote excitation. (**g**) Second order correlation function of the GeV emitter confirming a single photon emission.

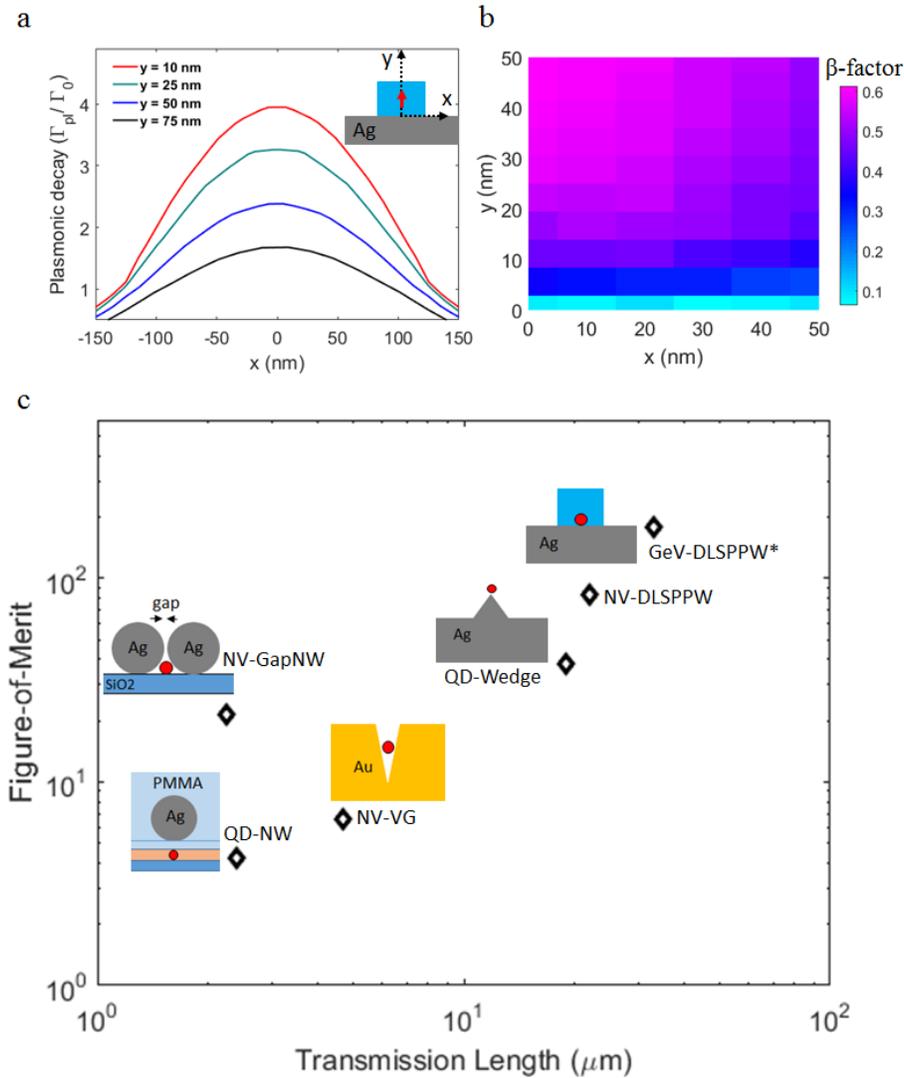

**Figure 5** (**a**) Simulated plasmonic decay rate ($\Gamma_{pl}/\Gamma_0$) for the DLSPPW coupled GeV center. Inset shows the cross section of a y-oriented dipole emitter inside the DLSPP waveguide (top right). (**b**) Distribution profile of the *β*-factor i.e., $\Gamma_{pl}/\Gamma_{tot}$, for a distribution of GeV-center inside a nanodiamond, where each colored square represents the central value of the corresponding in-plane dipole position. (**c**) Figure-of-merit (FOM) and transmission length of hybrid quantum plasmonic systems. The FOM of GeV-DLSPPW on Ag crystal is compared with other demonstrated quantum emitter-plasmonic waveguide (QE-PW) hybrid systems, including quantum dot-Ag nanowire (QD-NW)[6], NV-Gap Ag nanowire (NV-GapNW)[10], NV-V groove channel waveguides (NV-VG)[11], QD-Wedge waveguides (QD-wedge)[12], and NV-DLSPPW on Ag film[13]. The black diamond markers in the graph are extracted from the experimental results reported for the corresponding hybrid systems.

Supplementary Materials for

# On-chip excitation of single germanium-vacancies in nanodiamonds embedded in plasmonic waveguides

Hamidreza Siampour[*,1], Shailesh Kumar[1], Valery A. Davydov[2], Liudmila F. Kulikova[2], Viatcheslav N. Agafonov[3], and Sergey I. Bozhevolnyi[1]

[1] Centre for Nano Optics, University of Southern Denmark, Campusvej 55, Odense M, DK-5230, Denmark

[2] L.F. Vereshchagin Institute for High Pressure Physics, Russian Academy of Sciences, Troitsk, Moscow, 142190, Russia

[3] GREMAN, UMR CNRS CEA 6157, Université de Tours, 37200 Tours, France

[*]E-mail: hasa@mci.sdu.dk

## S1. GeV nanodiamond growth

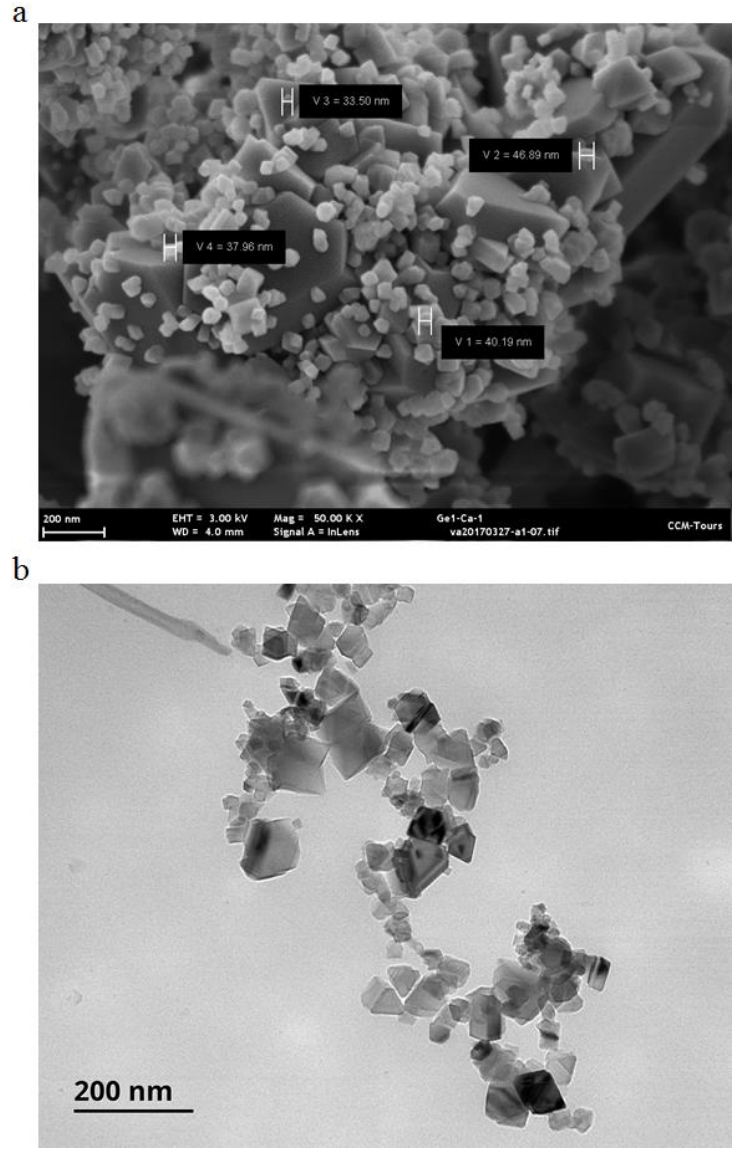

**Figure S1** (**a**) Scanning electron microscopy (SEM) image of the GeV nano and micro diamonds performed on the "raw" sample (after HPHT synthesis) (**b**) Transmission electron microscopy (TEM) image of the GeV nanodiamonds of different size taken after the chemical and ultrasonic treatment. Chemical treatment was carried out by three highly concentrated $HNO_3$ + $HClO_3$ + $H_2SO_4$ acids (at 200 °C for 3 hours) to remove traces of graphite. The ultrasonic treatment was done with a UP200H device (Hielscher).

## S2. Distribution of fluorescence lifetimes for GeV emitters in nanodiamonds

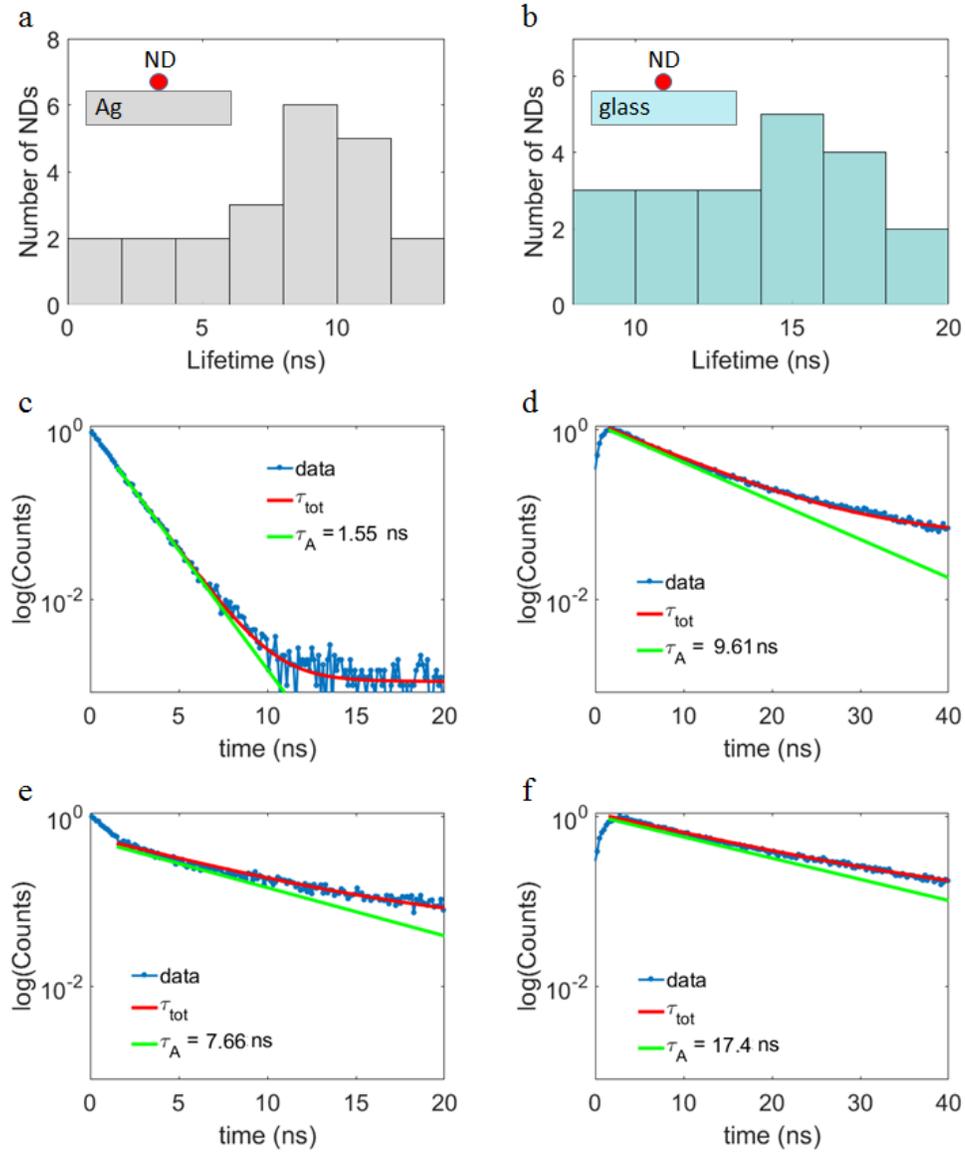

Figure S2 (**a**, **b**) Histogram graphs for lifetime distribution of GeV nanodiamonds on Ag surface (a), and on glass (b). Average lifetime of GeV nanodiamond on Ag plate is calculated to be 8.1. We have also calculate lifetimes for bare NDs (NDs on glass), where there is no influence from Ag plate and observed longer lifetimes (~15.9 ns on average), suggesting a 2-fold lifetime reduction due to the Ag plate. (**c**, **e**) Lifetime measurements on Ag plate for two cases in short range (c) and middle range of our collected data (e). $\tau_{tot}$ denotes total decay calculated from single exponential fit ($\tau_{tot}=A.\exp(-t/\tau_A)+c$, in which c is constant). (**d**, **f**) Lifetime measurements on glass for two cases in short range (d) and middle range of our collected data (f).

## S3. Propagation characteristics of DLSPPW on Ag crystals

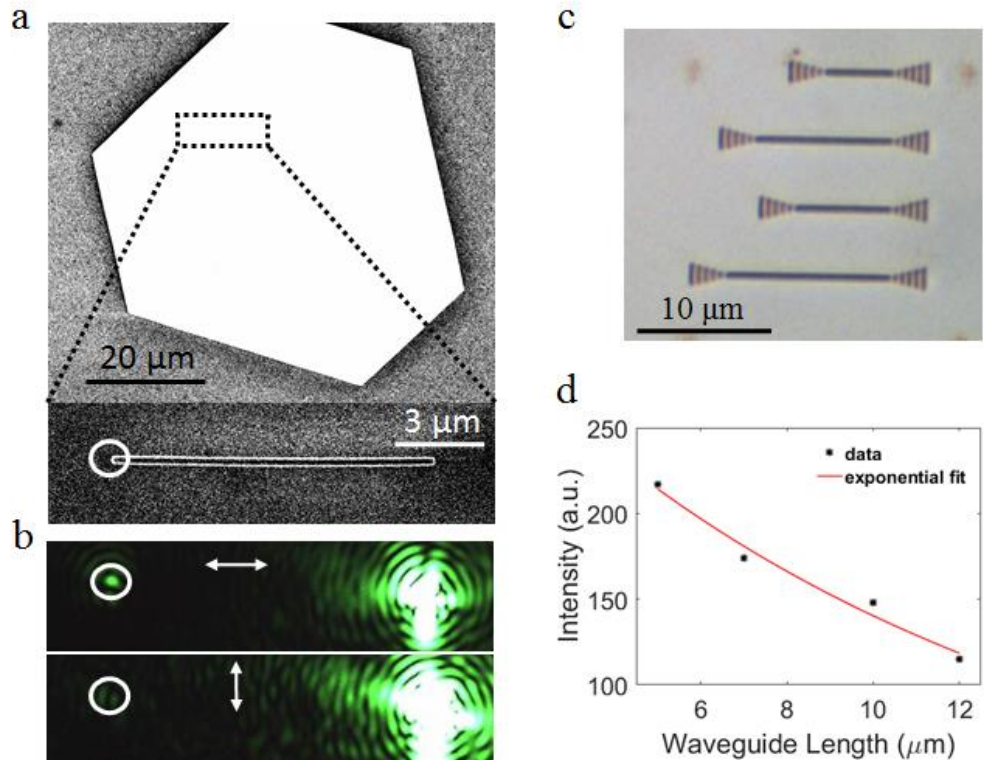

**Figure S3** (**a**) SEM image of a fabricated waveguide on Ag crystal flake. (**b**) Optical characterization of a DLSPPW without grating coupler. (**c**) Brightfield microscopy image of fabricated waveguides with grating coupler and different lengths. (**d**) Measured data for propagation characteristics of the DLSPPWs on Ag crystals with different lengths (black dot), and exponential fitting curve (red line), giving the propagation length of 11.8 μm at λ=532 nm (pump laser) for the DLSPPW on Ag flake.

## S4. Coupling efficiency of GeV-DLSPPW system

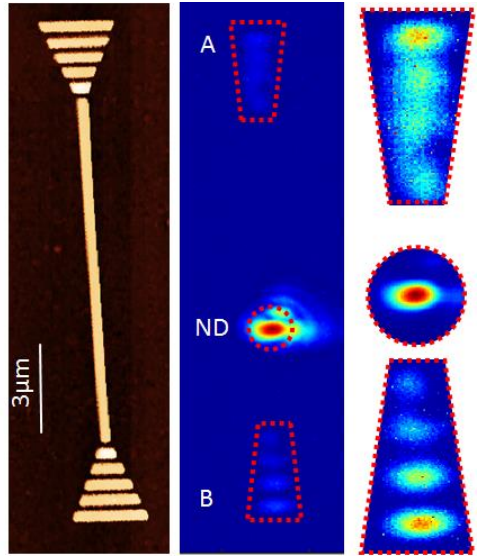

**Figure S4** AFM image of the fabricated waveguide on Ag flake (left), and CCD camera image of the whole structure where the nanodiamond is excited and a fluorescence image of the focal plane is taken (right). The 1/e propagation length, $L_P$, is extracted from the fluorescence signals at the two ends using $P_A/P_B = \exp[(L_A - L_B)/L_P]$, in which $L_A = 8$ μm and $L_B = 4$ μm. We assume symmetric coupling in two directions, uniform losses across the waveguide and the same out-coupling efficiency at the grating ends. The collected data are fitted to obtain the propagation length of 33 ± 3 μm for the GeV-DLSPPW hybrid system on Ag crystal flake that is even higher than the NV- DLSPPW system on Ag film, indicating a low material loss for the single crystalline Ag flakes.